\documentclass[lettersize,journal]{IEEEtran}

\usepackage{amsmath,amsfonts}
\usepackage{algorithmic}
\usepackage{algorithm}
\usepackage{array}
\usepackage[caption=false,font=normalsize,labelfont=sf,textfont=sf]{subfig}
\usepackage{textcomp}
\usepackage{stfloats}
\usepackage{url}
\usepackage{verbatim}
\usepackage{graphicx}
\usepackage{cite}
\usepackage[thinc]{esdiff}
\usepackage{siunitx}
\newcommand{\trevised}[1]{{\color{black}{#1}}}

\begin{document}

\title{Single Input Multi Output Model of Molecular Communication via Diffusion with Spheroidal Receivers}

\author{Ibrahim Isik, Mitra Rezaei, and Adam Noel,~\IEEEmembership{Member,~IEEE}

\thanks{Manuscript received 21 May 2024. This work was supported by the Engineering and Physical Sciences Research Council under Grant EP/V030493/1. (Corresponding author: Ibrahim Isik.)}
\thanks{The authors are with the School of Engineering, University of Warwick, CV4 7AL Coventry, U.K. (email: ibrahim.isik@warwick.ac.uk; Mitra.Rezaei@warwick.ac.uk;	Adam.Noel@warwick.ac.uk)}
\thanks{Ibrahim Isik is also with the Engineering Faculty, Inonu University, Malatya, Turkey.}
\IEEEpubid{0000--0000/00\$00.00~\copyright~2021 IEEE}}

\markboth{Isik \MakeLowercase{\textit{et al.}}: Single Input Multi Output Model of Molecular Communication}%
{Shell \MakeLowercase{\textit{et al.}}: Single Input Multi Output Model of Molecular Communication}


\maketitle

\begin{abstract}
Spheroids are aggregates of cells that can mimic the cellular organization often found in tissues. They are typically formed through the self-assembly of cells in a culture where there is a promotion of interactions and cell-to-cell communication. Spheroids can be created from various cell types, including cancer cells, stem cells, and primary cells, and they serve as valuable tools in biological research. \trevised{In this letter, molecule propagation from a point source is simulated in the presence of multiple spheroids to observe the impact of the spheroids on the spatial molecule distribution. The spheroids are modeled as porous media with a corresponding effective diffusion coefficient. System variations are considered with a higher spheroid porosity (i.e., with a higher effective diffusion coefficient) and with molecule uptake by the spheroid cells (approximated as a first-order degradation reaction while molecules diffuse within the spheroid). Results provide initial insights about the molecule propagation dynamics and their potential to model transport and drug delivery within crowded spheroid systems.}
\end{abstract}


\section{Introduction}

\IEEEPARstart{S}PHEROIDS are aggregates of cells that often adopt a spherical shape but are capable of mimicking tissue organization to some extent. They reflect the diversity in complexity and functionality within cultured systems, offering researchers different models suited for various experimental needs and objectives \cite{clevers2016modeling}. Spheroids serve as valuable tools to study cellular behavior, disease modeling, drug development, toxicity, and tissue engineering.

\trevised{Molecular communication (MC) is a bio-inspired paradigm to realize communication, particularly at micro- and nano-scales, using molecules as information carriers between transmitters (TXs) and receivers (RXs). MC can serve as a useful aid to study the propagation of molecules through spheroidal environments. Initial work to model spheroids with MC include \cite{schafer2022channel} and our own group's contributions in \cite{rezaei2023spheroidal, arjmandi20233d}. \cite{rezaei2023spheroidal} presented an end-to-end diffusive MC system with a spheroidal TX and spheroidal RX. The molecule concentrations inside and outside the RX were analytically derived and verified with particle-based simulations. The results showed that the porous structure of the receiving spheroid both amplified and dispersed diffusion signals, thus there was a trade-off between porosity and the potential information transmission rate.
	
A limitation in the analytical models for molecule propagation in these initial studies is that they considered individual spheroid RXs. In reality, spheroidal receivers can exist in populations alongside other spheroids of the same or different types, as discussed in \cite{clevers2016modeling}. This detail is particularly crucial as it reflects real scenarios such as organ-on-chip systems, where experimental wells could contain dozens of spheroids that signal each other \cite{shroff2022studying,adepu2021controlled}.

There is well-established literature in the MC community to consider multi-RX systems, e.g., \cite{sabu20203,yaylali2023channel,bao2019channel,kwak2020two,huang2020channel}. However, these contributions have focused on simpler RX models where the entire RX is a perfect molecule absorber, i.e., all molecules that make contact with the RX are permanently removed from the system, such that the distribution of molecules throughout an RX is irrelevant. Spheroids are much more complex RXs that \textit{effectively have the potential for both reversible and irreversible absorption}, since molecules can 1) diffuse both into and out of spheroids; \textit{and} 2) be uptaken by individual cells while diffusing inside the spheroid.

The nature of diffusion across the boundary between a spheroid and its external fluid medium was the focus of \cite{arjmandi20233d}. It was revealed that the boundary is characterized by an amplification factor and that factor is a function of the diffusion coefficients of the spheroid and the medium. The existence of the amplifying boundary condition is evident in natural systems such as tissues and tumors in multi-cellular organisms, as well as bio-films of microorganisms.

Of particular significance for local molecule uptake is its relevance for drug delivery; there is substantial interest in understanding drug penetration across spheroids -- which are often assembled from tumor cells -- in order to optimize dosing strategies for drug effectiveness. \cite{leedale2020multiscale} analyzed cell membrane permeation and porosity across a spheroid structure and unveiled the consequential effects of these properties on the penetration of drugs. Nanocarriers can facilitate drug delivery to sites challenged by anatomical barriers, where drug penetration is otherwise hindered. Therefore, employing nano-structures or nanocarriers to transport drugs can greatly enhance drug distribution throughout the body and contribute to maximizing therapeutic efficacy \cite{shroff2022studying}. In the context of drug delivery, a mathematical model was developed to describe the delivery and penetration of nanoparticle drug carriers into spheroids in the radial direction \cite{goodman2008spatio}. This model enabled predicting nanoparticle penetration into spheroids under various conditions and understanding the impact of tumor architecture on nanoparticle delivery efficiency.}
 

\trevised{It is therefore evident that there are interesting open challenges to model systems with multiple spheroidal RXs. In this letter, we seek to develop an understanding of diffusing molecule propagation in a multi-spheroid system with a single point TX and multiple spheroid RXs, i.e., a single input and multiple output (SIMO) spheroid system. We are particularly interested in the applicability of our group's contributions for the single-spheroid RX system in \cite{rezaei2023spheroidal}; i.e., can we identify scenarios where it would be appropriate to approximate the SIMO system as one with a single spheroid receiver? With a particle-based simulator (PBS) developed in MATLAB, we use color maps to visualize the time-varying spatial molecule profile of unbounded SIMO spheroid systems. Future work will develop corresponding analytical models. Herein, we focus on the impact of the number of spheroids and the spheroid porosity. As a special case designed for drug delivery modeling, we also consider the profile of absorbed molecules when individual cells within the spheroids are able to irreversibly uptake the diffusing molecules.}


The rest of the letter is organized as follows. Section II describes the system model and PBS of the spheroid SIMO system. Section III presents the results. Finally, Section IV concludes the letter.

\section{System Model}

\begin{figure}
	\centering
	\includegraphics[width=2.5in]{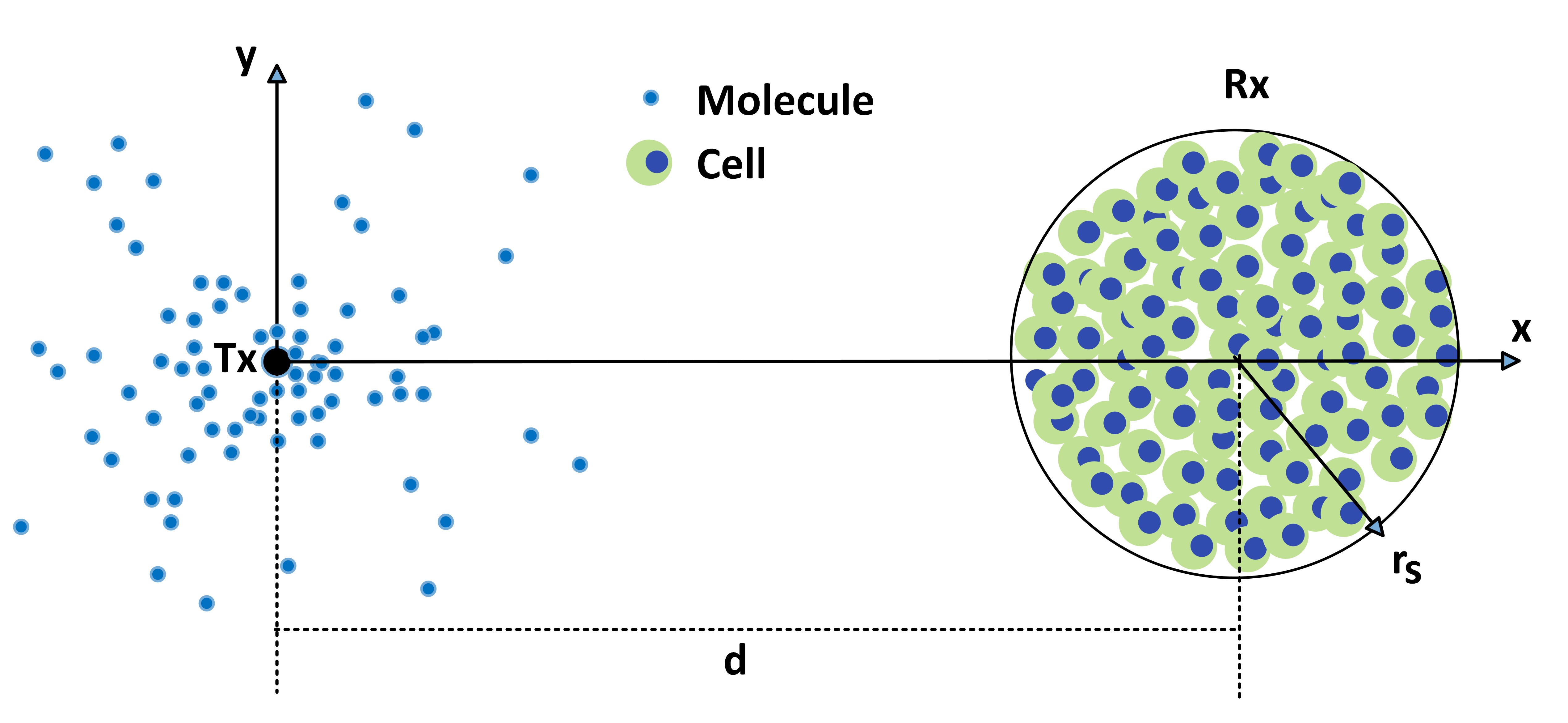}
	\caption{A schematic representation of 1 spheroid RX scenario with a point transmitter. The receiver is placed at a distance $d$ from the transmitter. Molecules are emitted from TX at $t$ = 0.}
	\label{fig_1}
\end{figure}

\trevised{We consider a diffusive MC system with a point TX and one or more spheroidal RXs in an unbounded fluid medium. The proposed model with \trevised{1 RX} is shown in Fig. \ref{fig_1}.} The point TX is positioned at the origin of a two-dimensional (2-D) Cartesian coordinate system (0, 0) and the \trevised{spheroid RX is centered at $(d, 0)$. We use a 2-D model in this letter so that we can clearly visualize the spatial molecule profiles inside and outside the spheroidal RXs. Nevertheless, the RX parameters are informed by 3-D spheroids and future model iterations will be in 3-D.} The transmitter emits signaling molecules of any type. The released molecules undergo random movement in the environment following independent Brownian motion.

\trevised{We model RXs as porous media comprised of a homogeneous distribution of cells. Molecules that enter an RX can diffuse through the extracellular space around the cells. We define the porosity parameter $\epsilon$ from a 3-D representation of spheroids as follows. Given total spheroid volume $V_s$ and a total of $N_c$ constituent cells (each of volume $V_c$) within a spheroid, the corresponding porosity is}  \cite[Eq.~(1)]{rezaei2023spheroidal},

\begin{equation}
\epsilon=\frac{V_s-N_cV_c}{V_s}.
\label{eq_1}
\end{equation}

\trevised{Over a given time interval, the porosity shortens the net displacement of molecules within a spheroid.} Consequently, the effective diffusion within a whole spheroid volume is reduced compared to the free fluid diffusion outside. The effective diffusion coefficient $D_\text{eff}$ within a spheroid can be determined from the free fluid diffusion coefficient $D$ outside the spheroid, as

\begin{equation}
D_\text{eff}=\frac{\epsilon}{\tau}D,
\label{eq_2}
\end{equation}
where $\tau$ is the tortuosity, which refers to the degree of path irregularity or curvature experienced by a molecule while it traverses through the extracellular space of a spheroid, and is modeled as  $\tau=\epsilon^{-0.5}$ \cite{sharifi2019numerical}.

\trevised{By modeling diffusion within a spheroid via $D_\text{eff}$, we implicitly assume that the cells therein are solid and reflecting. To enable molecule uptake by the cells, where the diffusing $A$ molecules are converted into \textit{non-diffusing} $E$ molecules, we consider a first-order approximation of this conversion via the reaction
	\begin{equation}
	A  \xrightarrow{k_f} E,
	\label{eq_1_1}
	\end{equation}
while molecules are within a spheroid, where $k_f$ is the first-order conversion rate.}

\begin{figure}[!t]
\centering
\includegraphics[width=2.5in]{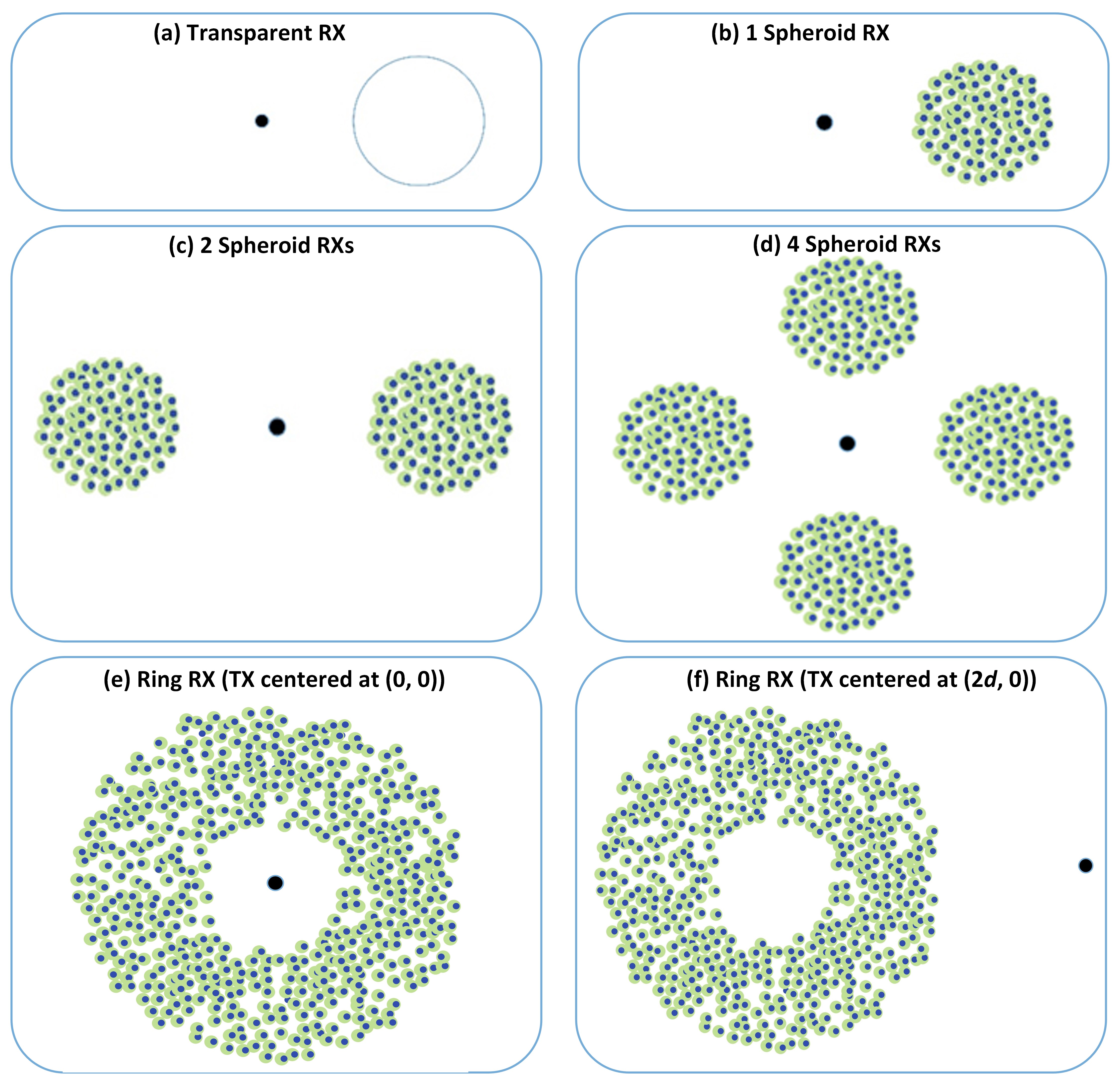}
\caption{The proposed SIMO model with a point transmitter and different \trevised{spheroid RX scenarios.} (a) 1 transparent RX; (b) 1 spheroid RX; (c) 2 spheroid RXs; (d) 4 spheroid RXs; (e) ring RX (TX at (0, 0)); and (f) ring RX (TX at (2$d$, 0)).}
\label{fig_2}
\end{figure}

\trevised{We generalize the single-spheroid RX system to consider all of the cases shown in Fig.~\ref{fig_2}. Fig.~\ref{fig_2}(a) is the standard transparent RX system for comparison, where the diffusion coefficient is $D$ everywhere (or, equivalently, where the spheroid has porosity $\epsilon=1$ and $D_\text{eff}=D$). Fig.~\ref{fig_2}(b) is the single-spheroid RX system as in Fig.~\ref{fig_1}. For symmetry and to facilitate comparison with Fig.~\ref{fig_2}(b), Fig.~\ref{fig_2}(c) has 2 spheroids centered at ($d$, 0) and ($-d$, 0). Similarly, Fig.~\ref{fig_2}(d) has 4 spheroids centered at ($d$, 0), ($-d$, 0), (0, $d$), and (0, $-d$).} Figs. \ref{fig_2}(e) and \ref{fig_2}(f) depict scenarios where we approximate a large number of surrounding spheroids that effectively form a 2-D ring around the system origin. The TX is kept at the origin in Fig.~\ref{fig_2}(e) and is placed at (2$d$, 0) in Fig.~\ref{fig_2}(f). \trevised{Fig.~\ref{fig_2}(f) can represent a large spheroid with a heterogeneity in the center; spheroids often have heterogeneous substructures \cite{clevers2016modeling}.}

\subsection{Particle-Based Simulator}

We use a particle-based simulator (PBS) that is implemented in MATLAB \trevised{to visualize the non-homogeneous diffusion (i.e., with diffusion coefficient $D$ in the bulk medium and effective diffusion coefficient $D_\text{eff}$ within the spheroids). The values of these coefficients depend on various factors such as the size and shape of the spheroids, the medium in which diffusion occurs, and the properties of the diffusing molecules. In experimental studies, the diffusion coefficient within specific spheroid models such as tumors is typically measured using fluorescence techniques and has been observed between $10^{-9}\, \si{m^2s^{-1}}$ and $10^{-11}\,\si{m^2s^{-1}}$ in the literature \cite{mukomoto2020oxygen,grimes2018oxygen}.} Molecules released by the point TX move independently in two-dimensional (2-D) space with time steps of $\Delta t$ = 0.5\,s until $t_\text{end}$. The displacement of a molecule within $\Delta t$ s is modeled as a Gaussian random variable with zero mean and variance of $2D\Delta t$ along each Cartesian dimension \trevised{(where $D$ is replaced with $D_\text{eff}$ inside a spheroid)}. When the displacement vector calculated for a molecule outside a spheroid crosses a spheroid boundary in a given time slot, then the length of the portion of the vector inside the spheroid is \trevised{scaled by the factor $\sqrt{\frac{D_\text{eff}}{D}}$ (see \cite{arjmandi20233d})}. Conversely, if the calculated displacement vector for a molecule inside the spheroid crosses the spheroid boundary in a time slot, then the length of the portion of the vector outside the spheroid is \trevised{scaled by the factor $\sqrt{\frac{D}{D_\text{eff}}}$}. To simulate the impulsive source, $N=10^7$ molecules are released at the transmitter location. \trevised{For $V_c$ = 3.14$\times10^{-15}\, \si{m^3}$\, and $N_c$ = 24000, which are the approximate volume of HepaRG spheroids and the number of cells therein, respectively \cite{bauer2017functional}, we have $\epsilon$ = 0.1349 and correspondingly $D_\text{eff} = 5\times10^{-11} \si{m^2s^{-1}}$}. These and all other PBS parameters are listed in Table \ref{tab:table1}. 

\begin{table}[!t]
\caption{Simulation parameters\label{tab:table1}}
\centering
\begin{tabular}{|l||l|}
\hline
Parameter & Value\\
\hline
Spheroid radius, $r_s$& 275\,$\mu$m\\
Distance, $d$ & 500\,$\mu$m\\
Number of cells in spheroid, $N_c$ & 24000\\
Cell volume $V_c$ & 3.14$\times10^{-15}\, \si{m^3}$\\
Number of released molecules, $N$ & $10^7$\\
Simulation end time, $t_\text{end}$ & 1 hour\\
Diffusion constant, $D$ & $10^{-9}\, \si{m^2s^{-1}}$\\
Effective diffusion constant, $D_\text{eff}$ & $5\times10^{-11} \,\si{m^2s^{-1}}$\\
Time step, $\Delta t$ & 0.5\,s\\
Porosity, $\epsilon$ & \{0.1349, 0.2791\}\\
\trevised{Conversion rate, $k_f$} & \{0, 0.01\}\,$\si{s^{-1}}$\\
\hline
\end{tabular}
\end{table}

\section{Results}

\trevised{All simulations are initiated by the emission of molecules from the TX at $\mathnormal{t}=0$. The study considers the six scenarios presented in Fig.~\ref{fig_2}. All results are plotted with color maps displaying the number of molecules per pixel over the entire system, where the x and y axes represent distances in micrometers ($\si{\mu m}$) along the x and y directions, respectively. The location of the point TX is shown with a white dot.}

\begin{figure}[!t]
	\centering
	\includegraphics[width=3.5in]{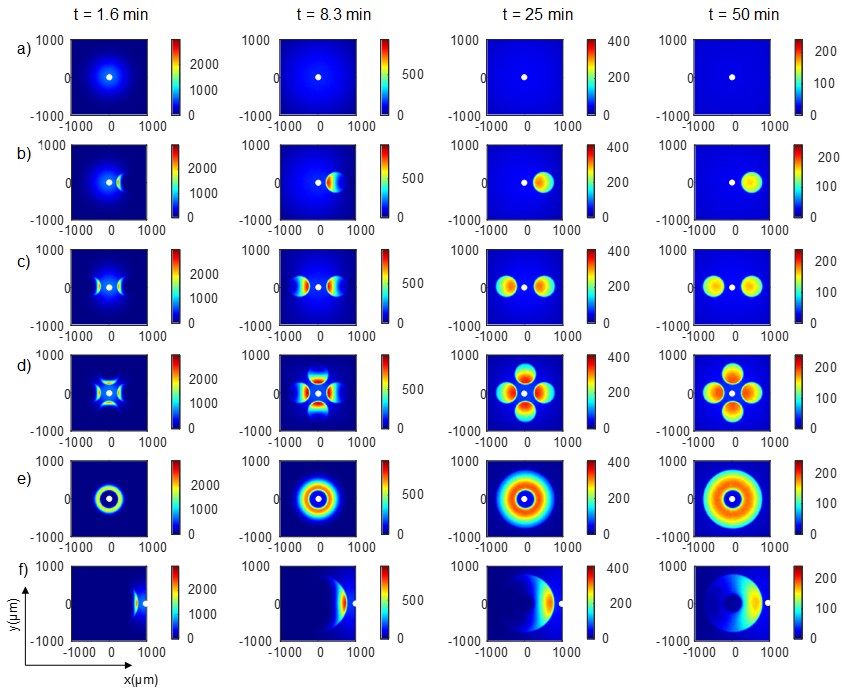}
	\caption{\trevised{Spatial molecule profile in 2D (x,y) for 6 different scenarios: (a) transparent; (b) 1 spheroid RX; (c) 2 spheroid RXs; (d) 4 spheroid RXs; (e) ring RX with TX in the middle; and (f) ring RX with the TX placed outside the RX at (1000,0) $\si{\mu m}$. The plots are shown from left to right for $\mathnormal{t}$ = \{1.6, 8.3, 25, 50\} min after transmission. The RX porosity is $\epsilon$ = 0.1349 and conversion rate is $k_f$ = 0. The colorbars show the diffusing molecule count for each pixel. Small white circles show the TX.}}
	\label{fig_3}
\end{figure}

The plots in Fig.~\ref{fig_3} depict the results for all six scenarios at times $\mathnormal{t} = \{1.6, 8.3, 25, 50\}$ min after transmission, progressing from left to right for $\epsilon$ = 0.1349 and $k_f$ = 0. As we expect, the initial \trevised{spatial molecule profile is very high around the TX in all cases due to the molecule release}. In the case of the ring RX in Fig.~\ref{fig_3}(e), many diffusing molecules received by the spheroid become entrapped within the ring structure. The porous nature of the ring RX, characterized by the lower diffusion constant \trevised{$D_\text{eff}$ for a large area around the TX}, poses a hindrance for these molecules to \trevised{either return to the center or diffuse to the edges of the region plotted. In contrast,} for the transparent scenario in Fig.~\ref{fig_3}(a), the concentration decreases over time across the entire environment because there is no spheroid (i.e., no porous environment) \trevised{to slow propagation down}, enabling molecules to diffuse away more easily due to the higher diffusion constant $D$.

\trevised{When comparing the scenarios in Fig.~\ref{fig_3}(b)-(d), i.e., with the different numbers of spheroid RXs, we initially see very similar spatial molecule profiles, i.e., at times $\mathnormal{t} = \{1.6, 8.3\}$ min, throughout each spheroid. This suggests that the initial impact of multiple local spheroids is minimal and a single-spheroid RX model such as that in \cite{rezaei2023spheroidal} would be sufficient to describe the propagation. However, the impact of additional spheroids becomes more apparent over time and actually leads to an \textit{increase} in the number of molecules observed within \textit{each} spheroid, particularly for the scenario with 4 spheroid RXs. This suggests that a local crowd of spheroids will eventually slow overall propagation such that molecules stay in the vicinity for a longer time, and a single-spheroid RX model would be insufficient.}

\trevised{We also comment on the scenario in Fig.~\ref{fig_3}(f), with the ring RX and the TX placed outside of it. We observe overall trends that are somewhat similar to the single spheroid RX in Fig.~\ref{fig_3}(b). However, by $\mathnormal{t}=50$ min, we can readily identify the inner region, and we observe that it maintains a lower spatial molecule profile than the surrounding ring. This scenario emphasizes the significance of heterogeneities in spheroid structures, and that internal areas described with higher diffusion coefficients will generally have fewer molecules present.}


\trevised{In Fig.~\ref{fig_4}, we consider the impact of increasing the spheroid porosity from $\epsilon$ = 0.1349 (as considered in Fig.~\ref{fig_3}) to $\epsilon$ = 0.2791, which reflects a decrease in the number $N_c$ of cells from 24000 to 20000. The conversion rate is kept at $k_f$ = 0. We then plot the spatial molecule profiles in the same manner as in Fig.~\ref{fig_3}, i.e., plotting at times $\mathnormal{t} = \{1.6, 8.3, 25, 50\}$ min after transmission, progressing from left to right, except that Fig.~\ref{fig_4} omits the transparent RX case (since this case does not depend on $N_c$).

\trevised{Most of the general trends observed in Fig.~\ref{fig_4} are consistent with those noted for Fig.~\ref{fig_3}; the ring RX surrounding the TX ``traps'' molecules, it takes time for local spheroids to have an impact on the molecules observed within a single spheroid, and with time we can identify the inner region when the TX is outside the ring. However, since these cases have a higher porosity, the molecules move faster and so the spatial profiles are more spread out and with lower molecule counts than in Fig.~\ref{fig_3}.}

	
\begin{figure}[!t]
	\centering
	\includegraphics[width=3.5in]{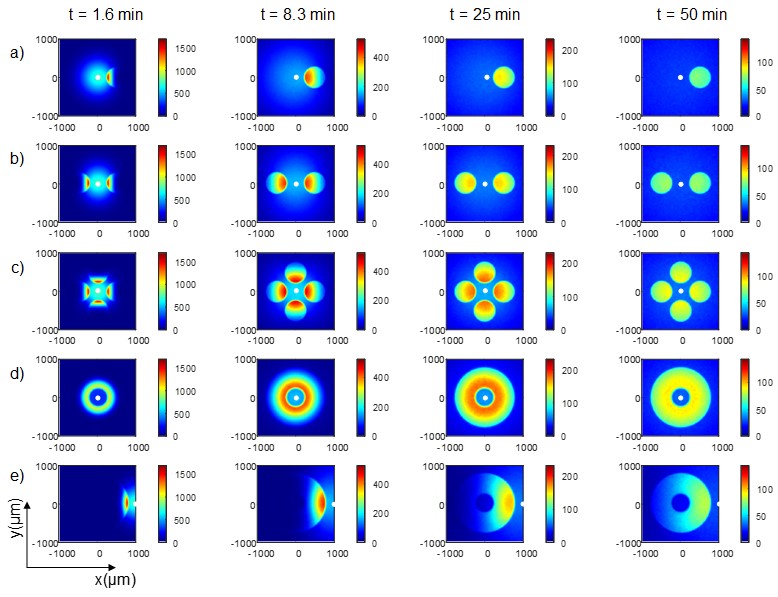}
	\caption{\trevised{Spatial molecule profile in 2D (x,y) for 5 different scenarios: (a) 1 spheroid RX; (b) 2 spheroid RXs; (c) 4 spheroid RXs; (d) ring RX with TX in the middle; and (e) ring RX with the TX placed outside the RX at (1000,0) $\si{\mu m}$. The plots are shown from left to right for $\mathnormal{t}$ = \{1.6, 8.3, 25, 50\} min after transmission. The RX porosity is $\epsilon$ = 0.2791 and conversion rate is $k_f$ = 0. The colorbars show the diffusing molecule count for each pixel. Small white circles show the TX.}}
	\label{fig_4}
\end{figure}

Finally, we consider the impact of a non-zero conversion rate within the spheroids, i.e., $k_f = 0.01\,\si{s^{-1}}$, in Fig.~\ref{fig_5}, and keep the porosity value at $\epsilon$ = 0.1349 (as in Fig.~\ref{fig_3}). The converted $E$ molecules are non-diffusing to approximate cellular uptake, so they accumulate within the spheroids over time. Fig.~\ref{fig_5} considers the scenarios with (a) 1 spheroid RX; (b) 2 spheroid RXs; and (c) 4 spheroid RXs, and the spatial profiles of the $E$ molecules are plotted at times $\mathnormal{t} = \{1.6, 8.3, 25, 50\}$ min after transmission, progressing from left to right. Since the $E$ molecules are non-diffusing, none are observed outside the spheroids.

\begin{figure}[!t]
	\centering
	\includegraphics[width=3.5in]{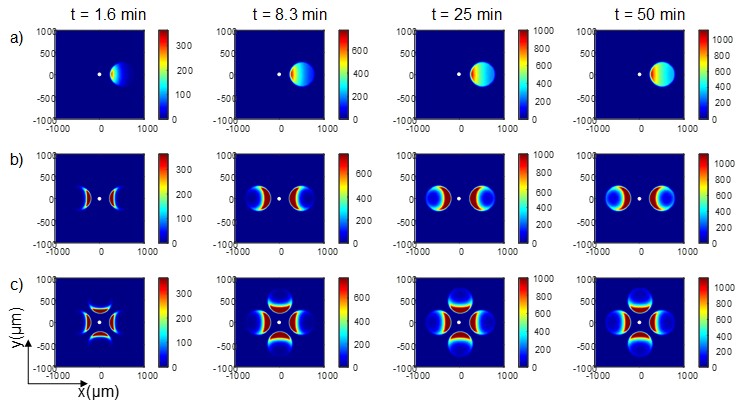}
	\caption{\trevised{Spatial molecule profile of converted molecules in 2D (x,y) for 3 different scenarios: (a) 1 spheroid RX; (b) 2 spheroid RXs; and (c) 4 spheroid RXs. The plots are shown from left to right for $\mathnormal{t}$ = \{1.6, 8.3, 25, 50\} min after transmission. The RX porosity is $\epsilon$ = 0.1349 and conversion rate is $k_f=0.01\,\si{s^{-1}}$. The colorbars show the non-diffusing converted molecule count for each pixel. Small white circles show the TX.}}
	\label{fig_5}
\end{figure}

There are a couple of very interesting observations to note from the spatial profiles of the non-diffusing $E$ molecules in Fig.~\ref{fig_5}. First, we see a much more significant difference between the 1-spheroid and 2-spheroid cases than what we observed in either Fig.~\ref{fig_3} or Fig.~\ref{fig_4}. For the case of 1 spheroid RX, there is a small molecule gradient throughout the spheroid that decreases away from the TX, suggesting that many molecules being converted were able to enter the spheroid at different angles. However, for the case of 2 spheroid RXs, the gradient in both of them is much larger, suggesting that a higher proportion of molecules being converted had entered the spheroid close to the TX. A similar trend is observed in the case of 4 spheroid RXs. Overall, the crowding due to multiple spheroids has a significant impact on the distribution of converted molecules, unlike the non-converting cases in Figs.~\ref{fig_3} and~\ref{fig_4}. This is not surprising, since the spheroids with a non-zero conversion rate are making permanent changes to the molecules that are converted.

Second, the spatial profile gradients of the non-diffusing $E$ molecules remain \textit{relatively stable} beyond the plot taken at $t=8.3$\,min, even though the number of molecules converted increases. We can see this by the similar patterns in the second, third, and fourth columns of Fig.~\ref{fig_5}, which are accompanied by increases in the colorbar scales as time progresses. This suggests that, for a given number of spheroids, the proportion of molecules entering the spheroid at a particular angle is consistent some time after the initial molecule release. Altogether, the results in Fig.~\ref{fig_5} emphasize the system sensitivity to molecule uptake, and which recall results from systems with multiple absorbing receivers as in \cite{yaylali2023channel}. However, unlike studies of conventional absorbing receivers, we have the additional insights due to the spatial molecule distributions within the spheroid RXs. For example, if the intent is to design or study a drug delivery system for cellular uptake, our model can inform how well a drug might propagate into a spheroid and how to optimize the dosage accordingly \cite{damrath2023optimization}.}


\section{Conclusion}

\trevised{This letter offered initial insights into the spatiotemporal distribution of molecules released into a system with one or more spheroidal receivers. Simulations were used to track diffusion and molecule uptake was approximated as a first-order conversion reaction with a non-diffusing product. The results demonstrated that, in the absence of uptake, the spatial molecule profile within a spheroid in the transient period is not substantially different when there are other spheroids nearby. This suggests that there is an interval for which a single-spheroid model (such as that in \cite{rezaei2023spheroidal}) could be sufficiently accurate. In the presence of uptake, the proximity of other spheroids makes a more significant difference when observing the number of molecules that have converted. The results enable us to develop intuition about the molecule propagation dynamics in crowded cellular systems, and can be helpful for applications that include communication and drug delivery. Future work will include studying the sensitivity to precise spheroid placement, extending the system model to 3-D, and deriving corresponding analytical results.}




\bibliographystyle{IEEEtran}
\bibliography{ref}

\end{document}